\documentstyle[12pt]{article}

\begin{document}
\thispagestyle{empty}
{\Large\bf
\begin{center}
Area-preserving structure of $2d$-gravity
\end{center}}
\bigskip
\bigskip
\bigskip
\bigskip
\bigskip
\bigskip
\begin{center}
{\large\bf
          D.R.Karakhanyan, R.P.Manvelyan, R.L.Mkrtchyan }{\footnote{E-mail:
mrl@dircom.erphy.armenia.su}}\\
\bigskip
{\it
     Yerevan Physics Institute,\\ Yerevan, Alikhanyan Br.st. 2,
     375036 Armenia}

\end{center}

\bigskip
\bigskip
\bigskip
\bigskip
     \begin{abstract} The effective  action  for  $2d$-gravity  with  manifest
area-preserving invariance is obtained in the flat and in  the  general
gravitational background. The  cocyclic  properties  of  the  last
action are proved, and generalizations on  higher  dimensions  are
discussed.
\end{abstract}
\setcounter{page}0
\newpage
\section{Introduction}

     The group of the area-preserving transformations  attracts  a
lot of attention recently, since it is a natural generalization of
a group of diffeomorphisms of a circle (the central  extension  of
algebra of which is the Virasoro algebra) and since it appears  in
many structures in modern $2d$ quantum field theory - in $W$-gravities
[1], matrix models [2] and particularly in $d=1$ string theory  [3],
where it appears as a group of hidden symmetry. The starting point
of our investigation is the attempt to develop an approach  to  $2d$
gravity, in which the  group  of  area-preserving  transformations
will play an important role just from the beginning. This  appears
to be possible - a room for such an approach i s  reserved  by  the
fact, that different regularizations,  generally,  have  an  equal
rights and lead to the same  physical  results.  Particularly,  we
note that if we apply a Weyl-invariant regularization procedure to
$2d$ gravity (instead of reparametrization-invariant one)  then  the
subgroup  of  surviving  invariance   from   the   group   of   $2d$
reparametrizations  will  be  a  group  of   the   area-preserving
transformations. In such an approach we have constructed the Weyl-
and area-preserving- invariant  effective  action  for  $2d$-gravity,
and, more generally,  found  a  rule  of  transformations  of  the
measure in the functional integral under reparametrization in such
a Weyl-invariant scheme of regularization. So, it  is  an  action,
integrated out not from  the  Weyl,  but  from  the  gravitational
anomaly,  and  it  possess  explicit  Weyl   and   area-preserving
invariances. In some  sense  our  approach  has  some  resemblance
with Polyakov's  light-cone  approach  [5],  in  which  a  residual
symmetry group is $SL(2,R)$, analog of which in our approach is  the
area-preserving  group.  The  exact  connection  with   Polyakov's
approach appears, when we use the area-preserving group to  fix  a
gauge $f^+(x)=x^+$  (in the parametrization $g_{\mu\nu}=
\rho\partial_\mu f^a\partial_\nu f^a$)   and  our
action coincides with Polyakov's one. So, in  some  sense  present
approach generalizes the Polyakov's  one  on  an  arbitrary  genus
surfaces.  We  also   make   contact   and   actually   obtain   a
generalization of Alekseev-Shatashvili's geometrical action [6].

  The other result of our approach is the construction  of  the
cocycle of the group of $2d$  diffeomorphisms,  which  seems  to  be
generalizable to higher dimensions.

     The paper is organized as follows. In Sect.2 we consider  the
general properties of gravitational  and  Weyl  anomalies  in  two
dimensions and construct the area-preserving effective action  for
$2d$ gravity.  In the Sect.3 we consider the problem of  the  change
of the measure under a general  diffeomorphism,  and  obtain  the
corresponding action, which is an action of Sect.2 in the  general
background metric.The cocyclic properties of that action  and  the
possible generalizations are discussed.

     \section{ The anomaly. }

     It  is  well-known,  that  quantization  of   two-dimensional
theory, invariant classically with  respect  to  reparametrization
and Weyl transformations  (e.g.  scalar  field,  interacting  with
gravitational one) leads to the unremovable anomaly, which usually
is presented as an anomaly of the Weyl invariance:
$$ \nabla_\mu T^\mu_\nu=0  \eqno (1) $$
$$ T^\mu_\mu=\frac{c}{24\pi}R  \eqno (2) $$
where $T^\mu_\nu$   is the energy-momentum tensor, $R$ is the scalar curvature
of the gravitational field $g_{\mu\nu}$  , and the  second  equation  express
the anomaly in the Weyl invariance, which  is  equivalent  to  the
tracelessness  of  $T^\mu_\nu$  .  One  can  make  a  redefinition  of   the
energy-momentum tensor
$$  \tilde T_{\mu\nu}=T_{\mu\nu}-(c/48\pi)g_{\mu\nu}R   \eqno       (3) $$
which satisfies now the equations  of  violated  reparametrization
invariance and non-violated Weyl invariance:
$$     \nabla_\mu\tilde T^\mu_\nu=-(c/48\pi)\partial_\nu R   \eqno    (4) $$

$$ \tilde T^\mu_\mu   = 0                     \eqno      (5) $$
The further correction of $T_{\mu\nu}$   by the metric-dependent terms allows
one to obtain  a  Weyl-covariant  formulae  for  reparametrization
anomaly:
     $$ \nabla_\mu T^\mu_\nu   = -(c/48\pi)\partial_\nu
(R(g_{\mu\nu}/\sqrt g)/\sqrt g) \eqno          (6) $$
where
 $$ R(g_{\mu\nu}/\sqrt g)=(R(g_{\mu\nu})-\Box\log\sqrt g)\sqrt g    \eqno
(7)$$ is now Weyl-invariant object. Difference of these forms of the same
re\-pa\-ra\-met\-ri\-za\-tion anomaly is a  reflection  of  a
 possibility  of
adding the local terms to the (effective)  action  to  shift  from
e.g. covariant to consistent form of the anomaly.  Explicitly,  we
present the following form of the effective action, which,  first,
differs from the standard non-local effective action
$$     S =\int  d^2x \sqrt g R \frac{1}{\Box}R \eqno       (8)  $$
by the local  terms,  and,  second,  is  Weyl-invariant  and  also
invariant w.r.t. the area-preserving transformations:
$$     S =\int d^2x \sqrt g R(g_{\mu\nu}/\sqrt g)\frac{1}{\sqrt g \Box}
\sqrt g R(g_{\mu\nu}/\sqrt g)     \eqno          (9)  $$
Introducing the parametrization of the metric  through  the  group
elements of the Weyl group and the group of diffeomorphisms:
$$     g_{\mu\nu}=\rho \partial_\mu f^a\partial_\nu f^a  \eqno  (10)  $$
we obtain for the effective action
$$     S_0=\frac{c}{48\pi} \int d^2x\frac{1}{\Delta^f_x}\{f^+,
\log\Delta^{f}_x \} \{f^- ,\log\Delta^f_x \}   \eqno      (11) $$
where
$$    \Delta^f_x=\{f^+ ,f^-\} =
\epsilon^{\mu\nu}\partial_\mu f^+\partial_\nu f^-  \eqno  (12)$$
The variation of $S$ under infinitesimal reparametrization
$x\to x+\epsilon (x)$ is
$$    \delta S_0=\frac{c}{96\pi} \int d^2x \partial_\mu \varepsilon^\mu
\{f^+,\{f^- ,\frac{1}{\Delta^f_x} \} \}=  $$

$$ =\int d^2x \partial_\mu \varepsilon^\mu \sqrt g (R(g_{\mu\nu} ) -
 \Box \log \sqrt g) \eqno (13) $$
in agreement with  (6).  Also,  from  (13)  we  see,  that $ S_0 $   is
invariant under area-preserving transformations, since for such  a
transformations $\partial_\mu\varepsilon^\mu = 0$.
The Weyl invariance of the action $S_0$   is
evident, since (11) doesn't depend on $\rho$. As was mentioned  in  the
Introduction, one can make contact with Polyakov's action  [4]  by
choosing the following gauge condition
$$     f^+ (x) = x^+    \eqno    (14)$$
and $ f$  appears to be the Polyakov's $f$ function.
It is well-known, that the transformation $f\to F$ through the relation
$$     F(t,f) = x         \eqno           (15)$$
(which means, that function $F$ is inverse to $f$ as a function of one
argument $ x$) leads to another form of the same action,  derived  by
the Shatashvili and Alekseev in  a  purely  geometrical  coadjoint
orbit approach - as a geometrical (i.e. with  initial  Hamiltonian
equal to  zero)  action  on  a  special  orbit  in  the  coadjoint
representation of Virasoro algebra  (the  generalization  of  that
approach on the group of the area-preserving  diffeomorphisms  see
in [5]). The natural generalization of (15) on  the  case  of  two
functions from two arguments leads  to  the  introduction  of  new
fields $F^+ ,F^-$  defined as an inverse to $f^+ ,f^-$ :
$$     F^\pm (f^+ ,f^- ) = x^\pm \eqno             (16)$$
In these new variables the action (11) looks as
$$     S_{geom}=\frac{c}{48\pi}\int
d^2x\partial_+\log\Delta^F_x\partial_{-}\log\Delta^F_x \eqno            (17)
$$ and is   invariant   with   respect   to   the   area-preserving
transformations, acting on the functions $F$:  $$     \delta F^\pm  = \pm
(\partial \omega (F)/\partial F^\pm)  \eqno (18) $$ and chiral
reparametrization $x^\pm\to x'^\pm (x^\pm )$.  So, action (17) have to be
considered as a generalization of the action  of  Shatashvili  and  Alekseev,
     and  the  interesting question is if  it  can  be  obtained  in  a
similar  geometrical fashion. The possible answer may  be  that  it  is  a
geometrical action on the $Diff_2 /SDiff_2$  orbit.

\section{ The transformation of the measure and cocycle properties.}

     Let's consider now the problem of the change of  the  measure
under the diffeomorphism $x\to f(x)$, which is closely connected to the
preceding considerations. Namely, under a diffeomorphism $f$
$$     D_{f^* g} \varphi = D_g\varphi e^{S(f,g)}  \eqno        (19) $$
This type of relation is very important, being the starting  point
of David, Distler and Kavai calculation of the critical  exponents
of $2d$ gravity [6].
     The action $S(f;g)$ have to satisfy some conditions. First,  in
the case of flat (or conformally flat) metric $g_{\mu\nu}$
   it  has  to  be equal to the previously derived action $S$ (11).
 Second, in the  case
of the infinitesimal diffeomorphism $f=x+\varepsilon$ it have to reproduce
Eq.(6) (or 13), namely:
$$     S(x+\varepsilon;g) = S(x;g) +
\frac{c}{96\pi}\int d^2x\partial_\mu\varepsilon^{\mu}\sqrt g
 (R(g_{\mu\nu}) -\Box \log\sqrt g) \eqno (20)$$
Finally, $S(f;g)$ have to  satisfy  the  following  property,  which
follows from the application of (19) to  the  composition  of  two
diffeomorphisms $f$ and $h$:
$$     S(hf;g) = S(f;h^{*}g) + S(h;g)  \eqno        (21)$$
which means, that $S(f;g)$  is  the  one-cocycle  of  the  group  of
two-dimensional   diffeomorphisms,   see   below.   Using    these
properties, one can find the following action:
$$     S(f;g) =\frac{c}{48\pi}\int d^2x\sqrt g[\frac{1}{2} g^{\mu\nu}
\partial_\mu \log\Delta^F_x\partial_\nu \log\Delta^F_x$$
$$+(R(g_{\mu\nu})-\Box \log\sqrt g)\log\Delta^F_x] \eqno (22)$$
where functions $F$ are inverse to $f$ (see (16).
     The  property  (21)  has  the  following  group-cohomological
interpretation  [9].  Let's  introduce  a  space  of   the   local
functionals $S(f_1,...,f_n ;g)$ of the $n$  elements  of  the  group  of
diffeomorphisms (i.e. $n$  pair  of  functions  $f^\mu_1,...,f^\mu_n)$
 and  the metric $g$  , and introduce the following  $d$  operation  which
acts from the space of the $S_n$  functionals to  the  space  of  the
$S_{n+1}$ functionals:
$$    \delta S_0 (f;g) = S_0(f^{*}g)-S_0(g)     \eqno    (23a)$$
$$    \delta S_1 (f_1,f_2;g) = S_1(f_1;g) + S_1(f_1;f_2^{*}g) -
 S_1(f_1 f_2 ;g)  \eqno (23b)$$
     etc. ([8])
with the standard property $\delta^{2}= 0$. As usual, the functions $S_n$  with
the property $\delta S_n =0$, modulo functions of the type $\delta S_{n-1}$   ,
 form  the
space of non-trivial cohomologies. In our case, if we consider the
space of local, Weyl-invariant functions, comparison of  (21)  and
(23b) shows, that $S(f;g)$ is the one-cocycle.  Evidently,  in  the
same way one can try to derive the formulae for the 1-cocycles  of
the group of the diffeomorphisms in  higher  dimensions,  starting
from the  theories  with  anomalous  Weyl  invariance  with  known
answer for the anomaly. Derivation of the corresponding  formulae
is in progress.
     The further natural directions of  investigation,  using  the
action (21) are: application of DDK approach [7] and the study  of
possible connection of the  present  approach  to  area-preserving
structures, found in [4] at $c=1$.
\section{Acknowledgments}
Authors would like to thank  A.Sedrakyan  and  O.Khudaverdyan
for helpfull discussions.

This work was supported in part by the grant 211-5291 YPI  of
the  German  Bundesministerium  fur  Forschnung  und  Technologie,
Federal Republic of Germany.
 

\begin{thebibliography}{99}
\bibitem{1} C.M.Hull, Nucl.Phys. B353 (1991) 707
\bibitem{1} Fukuma M., Kawai H. and Nakayama R.,  Int.  J.  Mod.  Phys.  A6
        (1991) 1385
\bibitem{1} J.Avan and A.Jevicki Mod.Phys.Lett. A7 (1992) 357
\bibitem{1} E.Witten and B.Zweibach Nucl.Phys. B377 (1992) 55 \\
       E.Witten Nucl.Phys. B373 (1992) 187
\bibitem{1} A.M.Polyakov, Mod.Phys.Lett. A2 (1987) 893
\\        V.G.Knizhnik, A.M.Polyakov, A.B.Zamolodchikov
Mod.Phys.Lett. A3  (1988) 819
\bibitem{1} A.Alekseev and S.Shatashvili, Nucl.Phys. B323 (1989) 719
\bibitem{1} F.David, Mod.Phys.Lett. A3 (1988) 1651
\\J.Distler, H.Kawai, Nucl.Phys. B321 (1989) 509
\bibitem{1} H.Aratyn, E.Nissimov, S.Pacheva  and  A.H.Zimerman,
Phys.Lett.  B242(1990)377
\\  R.P.Manvelyan and R.L.Mkrtchyan, Phys.Lett. B311 (1993) 51
\bibitem{1} Faddeev L.D. and Shatashvili S. Teor.Mat.Fiz. 60 (1984) 206
\end{thebibliography}
\end{document}